\documentclass{article}
\usepackage{spconf,amsmath,graphicx,hyperref,amssymb,booktabs,tabularx,array,multirow}

\title{DUPAR: Dual-Path Conversational Retrieval via Speech Retriever \\ with Cross-Turn Evidence Caching}
\vspace{-20pt}

\name{\shortstack{
Yuanjun Li$^{1,2,\dagger}$, Yiwen Liu$^{2,\dagger}$, Dapeng Li$^{2,3,\dagger}$,\\
Zhiwei Xu$^{1}$, Bin Zhang$^{4}$, Shengtao Zhang$^{2,*}$, Rong Shen$^{2,*}$
}\thanks{$^{\dagger}$Equal contribution.
$^{*}$Corresponding authors.}}
\address{
$^{1}$ Shandong University \quad $^{2}$ Li Auto Inc. \quad $^{3}$ Tsinghua University\\
$^{4}$ National Key Laboratory of Cognition and Decision Intelligence for Complex Systems,\\Institution of Automation, Chinese Academy of Sciences\\
liyuanjun@mail.sdu.edu.cn
}

\begin{document}
%
\maketitle
%

\begin{abstract}
Voice assistants grounded in external knowledge typically use automatic speech recognition (ASR) to transcribe speech queries before retrieving evidence from textual knowledge bases.
This cascade adds latency and propagates recognition errors, whereas direct speech retrieval is vulnerable to cross-modal misalignment.
To address these limitations, we propose \textbf{DUPAR}, a conversational retrieval framework with complementary slow and fast paths.
The fast path uses a task-adapted audio encoder aligned with frozen BGE-M3 text embeddings to search a cross-turn evidence cache.
When cache confidence is insufficient, the slow path fuses full-index retrieval using audio and ASR-transcript embeddings, and the selected evidence refreshes the next-turn evidence cache through one-hop graph expansion.
On a domain-specific knowledge base, our trained audio encoder approaches text-retrieval accuracy on clean speech with a $\boldsymbol{3.75\times}$ \textbf{query-side speedup} over ASR + Text Encoder.
It raises average Recall@10 from 0.771 to 0.875 on the noise benchmark and improves overall Recall@1 by 4.2 percentage points across synthesized speaking styles.
Compared with full-index audio retrieval, cross-turn evidence caching significantly reduces retrieval errors when the previous turn retrieves correct evidence and the follow-up targets a one-hop neighboring chunk.
\end{abstract}

\section{Introduction}
\label{sec:introduction}

\begin{figure}[!t]
    \centering
    \includegraphics[width=0.9\columnwidth]{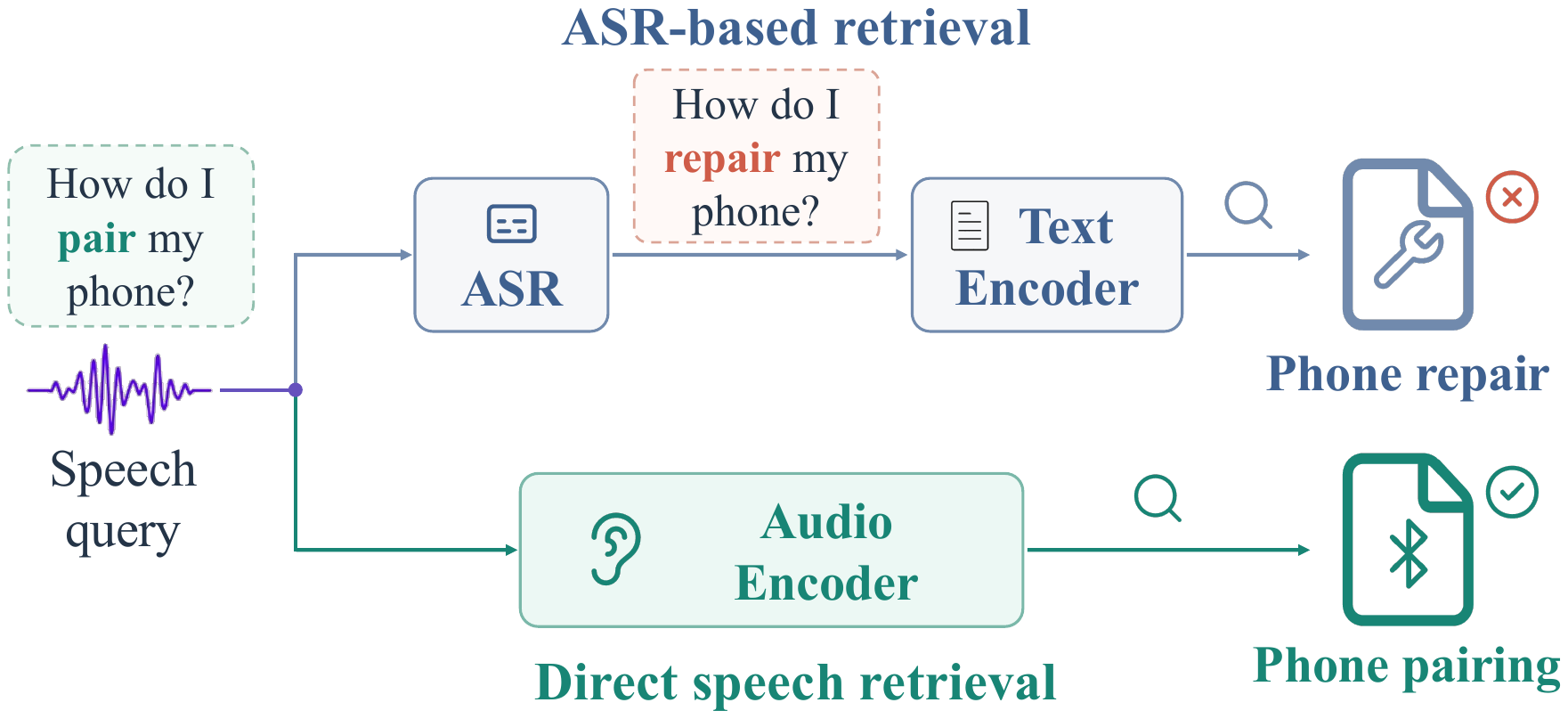}
    \vspace{-10pt}
    \caption{Illustrative comparison of retrieval paths. An ASR error changes ``pair'' to ``repair'', whereas direct speech retrieval bypasses transcription and its error propagation.}
    \label{fig:head}
    \vspace{-18pt}
\end{figure}

Knowledge-grounded voice assistants answer speech queries with evidence from domain-specific textual knowledge bases. Low response latency is essential because long delays disrupt conversational flow and degrade user experience~\cite{li2022speak,maslych2025response}. Retrieval-augmented generation (RAG) grounds responses in external knowledge that can be updated independently of model parameters~\cite{lewis2020rag}. However, speech queries and textual evidence create a modality mismatch. A common solution transcribes speech with automatic speech recognition (ASR) before text retrieval~\cite{thulke2024documentgrounded}. This cascade delays retrieval and can propagate recognition errors into downstream ranking~\cite{sidiropoulos2022impact,ravichander2021noiseqa}, as illustrated in Fig.~\ref{fig:head}. Rare domain-specific terms, including proper nouns, acronyms, and technical terminology, are especially vulnerable to mistranscription and may yield irrelevant or empty evidence~\cite{wang2020asrerror,suh2024domain}.

\begin{figure*}[!t]
    \centering
    \includegraphics[width=\textwidth]{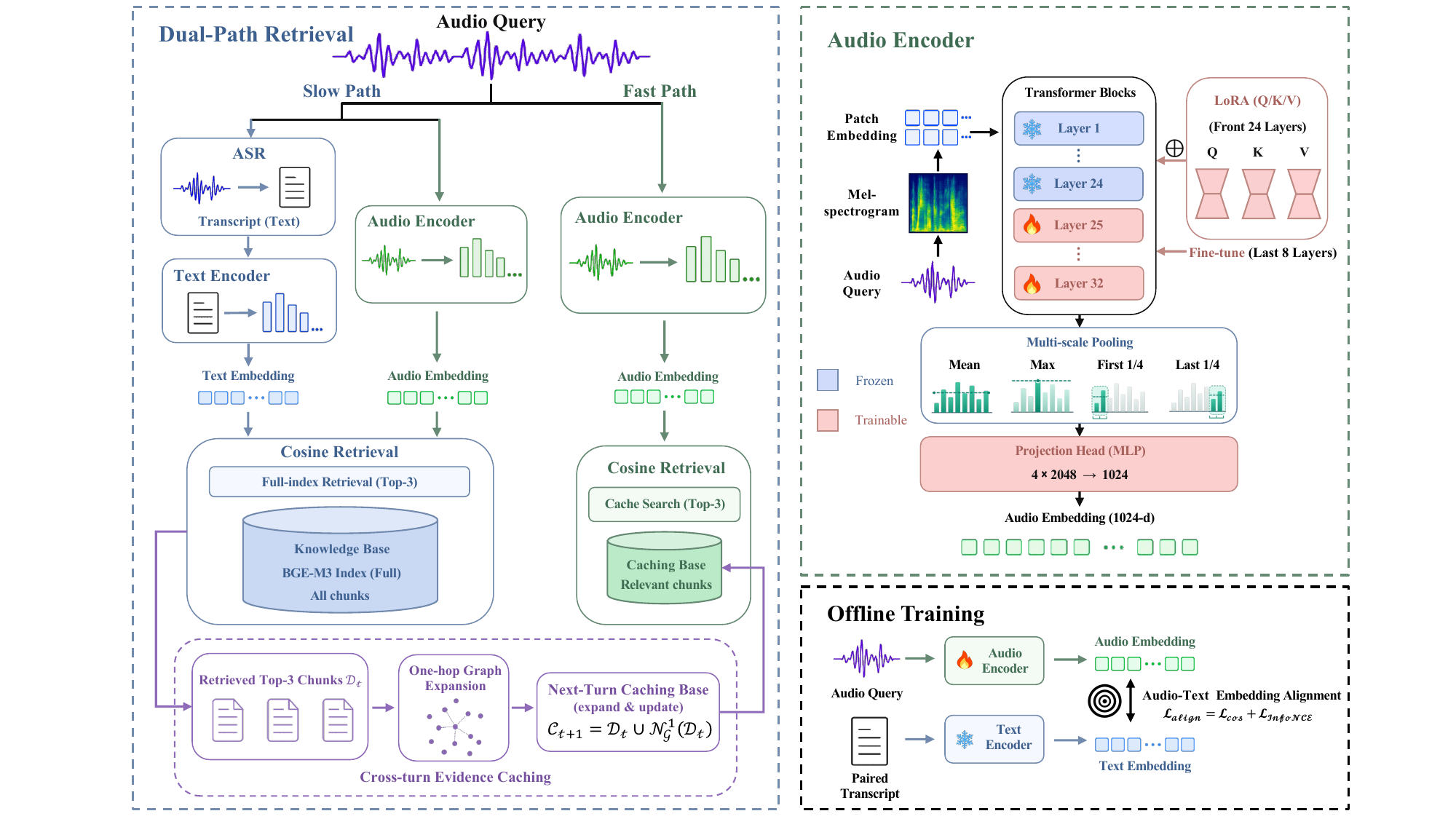}
    \vspace{-20pt}
    \caption{Overview of DUPAR. The fast path directly encodes the speech query and searches the cross-turn evidence cache, while the slow path fuses full-index retrieval using audio and ASR-transcript embeddings. The selected evidence and its one-hop neighbors refresh the cache for the next turn. Right: the audio encoder architecture and offline speech--text alignment.}
    \label{fig:method}
    \vspace{-10pt}
\end{figure*}

ASR-free systems instead encode speech queries and text chunks into a shared embedding space, avoiding an intermediate transcript~\cite{wang2024reslm,feng2025enhancing}. However, contrastive alignment can leave a residual modality gap~\cite{sofer2025modalitygap}, while acoustic noise can degrade retrieval from speech queries~\cite{li2026squtr}. When searching the full index, even modest misalignment may rank an unrelated chunk above the relevant evidence. In multi-turn conversation, the current query often depends on preceding turns~\cite{mo2024historyaware}. This dependency motivates using recently retrieved evidence as a prior for the next retrieval.

Motivated by these observations, we propose \textbf{DU}al-\textbf{PA}th conversational \textbf{R}etrieval (\textbf{DUPAR}). At each turn, the fast path maps the speech query to an audio embedding aligned with text embeddings from a frozen BGE-M3 encoder~\cite{chen2024m3embedding}. It searches the evidence cache constructed after the previous turn, while ASR runs in parallel. If the cache is nonempty and the highest similarity exceeds a confidence threshold, the fast result is committed immediately. Otherwise, the slow path produces two full-index rankings using the audio embedding and the text embedding of the ASR transcript, then fuses them. The selected evidence grounds the current response and seeds one-hop expansion over a static chunk relation graph, producing the evidence cache for the next turn.

Our contributions are as follows:
\begin{itemize}
    \setlength{\topsep}{2pt}
    \setlength{\partopsep}{0pt}
    \setlength{\itemsep}{0pt}
    \setlength{\parsep}{0pt}
    \vspace{-5pt}
    \item We propose a dual-path conversational retrieval framework. The fast path searches an evidence cache directly from speech, the slow path provides full-index fallback, and one-hop expansion of selected evidence updates the cache for the next turn.
    \item We train an audio encoder that aligns audio embeddings with those produced by a frozen text encoder, enabling direct reuse of precomputed chunk embeddings.
    \item Experiments show that our speech retriever achieves performance close to text retrieval while reducing query-side latency relative to ASR-based cascaded retrieval. It remains robust to synthetic noise and speaking style.
\end{itemize}

\section{Related Work}
\label{sec:related_work}

\noindent\textbf{ASR-based cascaded retrieval.} Prior work uses simulated ASR errors to train self-correction and knowledge-cluster prediction modules before reranking candidate documents~\cite{tam2022robust}. AVATAR transcribes speech and autoregressively generates document identifiers, using augmentation and contrastive learning for robustness~\cite{wang2023avatar}. CLKS contrasts aggregated ASR N-best histories with written-dialogue representations to select FAQ snippets~\cite{zhu2023contrastive}. Yet all remain transcript-dependent, so ASR errors can still propagate and retrieval must wait for transcription. Our trained audio encoder directly embeds the speech query, bypassing this error channel on the fast path and reducing query-side latency.

\noindent\textbf{Cross-modal retrieval.} SpeechRAG trains a speech adapter to match transcript embeddings from a frozen text retriever, enabling text queries to retrieve speech chunks~\cite{min2025speechrag}. WavRAG learns a unified retriever for hybrid text--audio knowledge bases~\cite{chen2025wavrag}. E2E RAG maps speech queries and text chunks into a shared embedding space for direct full-index retrieval~\cite{feng2025enhancing}. Like SpeechRAG, DUPAR aligns audio and text embeddings using a frozen text encoder, but retrieves text chunks from speech queries. It further combines cross-turn evidence caching with fused full-index fallback.

\noindent\textbf{Conversational retrieval.} ChatRetriever encodes the complete text session as a single dense representation~\cite{mao2024chatretriever}. VoiceAgentRAG predicts several follow-up topics with an LLM and issues full-index vector-store searches for these predictions in the background~\cite{qiu2026voiceagentrag}. This moves retrieval off the foreground path but introduces additional prediction, embedding, and retrieval overhead. We instead construct the next-turn cache from evidence already retrieved in the current turn and its one-hop graph neighbors, avoiding both future-topic prediction and speculative full-index searches.

\section{Method}
\label{sec:method}

\subsection{Dual-Path Retrieval and Evidence Caching}
\label{subsec:retrieval}

DUPAR (Fig.~\ref{fig:method}) represents each text chunk $d$ in the full knowledge base $\mathcal{K}$ with a fixed, normalized BGE-M3 embedding $\mathbf{e}_d$. At turn $t$, the fast path encodes speech query $a_t$ with the audio encoder $f_\theta$. It retrieves the top-$k$ chunks $\mathcal{D}_t^F$ by cosine similarity from the previous-turn evidence cache $\mathcal{C}_t$ while ASR runs in parallel. For a nonempty cache, confidence is $q_t^F=\max_{d\in\mathcal{C}_t}\cos(f_\theta(a_t),\mathbf{e}_d)$.

The slow fallback merges full-index rankings from $f_\theta(a_t)$ and the BGE-M3 embedding of the history-conditioned ASR transcript. It retains each chunk's higher score and selects the top-$k$ unique chunks $\mathcal{D}_t^S$. With a fixed confidence threshold $\gamma$, evidence is selected as
\begin{equation}
\mathcal{D}_t=
\begin{cases}
\mathcal{D}_t^F, & \mathcal{C}_t\neq\varnothing\ \text{and}\ q_t^F>\gamma,\\
\mathcal{D}_t^S, & \text{otherwise}.
\end{cases}
\label{eq:routing}
\end{equation}
Initially, $\mathcal{C}_1=\varnothing$, so the first turn is slow. After a fast selection, ASR triggers no further retrieval or response.
Selected evidence grounds the response and refreshes the cache through a static chunk relation graph $G$:
\begin{equation}
\mathcal{C}_{t+1}=\mathcal{D}_t\cup\mathcal{N}_G^1(\mathcal{D}_t).
\label{eq:cache_update}
\end{equation}
Here, $\mathcal{N}_G^1(\mathcal{D}_t)$ contains their one-hop neighbors. Graph edges encode embedding proximity and shared domain metadata.

\subsection{Audio Encoder}
\label{subsec:encoder}

We initialize a reconstructed encoder with Qwen3-Omni audio weights~\cite{xu2025qwen3omni}. A convolutional front end and a 32-layer Transformer map normalized log-Mel features to frame representations $H_t$. The first 24 backbone layers remain frozen, with low-rank adaptation (LoRA)~\cite{hu2022lora} added to their query, key, and value projections. The last eight layers are fully fine-tuned. Other pretrained weights remain frozen. Multi-scale pooling produces $\mathbf{z}_t^a=f_\theta(a_t)$:
\begin{equation}
\mathbf{z}_t^a=\operatorname{Norm}\!\bigl(\operatorname{LN}(W[\bar H_t;\max H_t;\bar H_t^{(1)};\bar H_t^{(4)}]+\mathbf b)\bigr).
\label{eq:pooling}
\end{equation}
$\bar H_t$ and $\max H_t$ denote global temporal mean and max pooling; $\bar H_t^{(1)}$ and $\bar H_t^{(4)}$ average the first and last quarters. The projection $W$ and bias $\mathbf b$ produce 1024 dimensions; $\operatorname{LN}(\cdot)$ and $\operatorname{Norm}(\cdot)$ apply layer normalization and $\ell_2$ normalization.

Frozen BGE-M3 provides target embeddings $\mathbf{z}_i^t$ from paired reference query texts, not ASR transcripts. We train audio embeddings $\mathbf{z}_i^a$ with
\begin{equation}
\mathcal{L}=\frac{1}{B}\sum_{i=1}^{B}w_i\!\left[1-\cos(\mathbf{z}_i^a,\mathbf{z}_i^t)\right]+\operatorname{CE}\!\left(Z^aZ^{t\top}/\tau,\mathbf{y}\right),
\label{eq:alignment}
\end{equation}
where $B$ is the batch size, $w_i$ upweights vague queries, and $Z^a,Z^t$ stack normalized audio/text embeddings. Cross-entropy $\operatorname{CE}(\cdot)$ implements audio-to-text in-batch InfoNCE with diagonal pairing labels $\mathbf y$ and temperature $\tau$.

\section{Experimental Setup}
\label{sec:experiments}

\noindent\textbf{Data and metrics.}
Our knowledge base contains 1,890 text chunks. CosyVoice3 text-to-speech (TTS) synthesis yields 30,215 utterances, split by utterance into 21,150/4,532/4,533 training/development/test samples. Clean evaluation uses 1,000 test utterances. Noise evaluation uses 500 queries under clean conditions and signal-to-noise ratios (SNRs) of 20, 15, 10, and 5\,dB, yielding 2,500 inputs. Speaking-style evaluation uses a separate 1,000-utterance TTS cohort with ASR outputs, covering 284 query clusters and four TTS configurations. We report Recall@$K$ ($K\in\{1,5,10,20\}$) and MRR@20 for retrieval coverage and ranking quality, respectively; noise evaluation uses Recall@10.

\noindent\textbf{Systems and training.}
The baseline, \emph{ASR + Text Encoder}, combines Qwen3-Omni ASR with BGE-M3 text encoding. \emph{Audio Encoder} directly embeds speech without transcription. Both systems perform retrieval over the same document embeddings. Training uses four A100 GPUs and 48 examples per GPU. AdamW learning rates are $2\times10^{-4}$ for LoRA and the pooling head, and $5\times10^{-5}$ for fully tuned layers. We select the best development checkpoint.

\noindent\textbf{Efficiency protocol.}
Query-side latency excludes recording and generation. A separate streaming benchmark measures time to first text (TTFT) and time to first audio (TTFA) from complete speech availability. These timings include generation up to the first output. We use 30 paired utterances on an isolated endpoint after three warm-up pairs. Added parameters count the text encoder for \emph{ASR + Text Encoder} and LoRA plus the pooling head for \emph{Audio Encoder}, excluding ASR and the existing audio backbone, including its fine-tuned layers.

\section{Results and Analysis}
\label{sec:results}

\subsection{Main Results}
\label{subsec:main_results}

In Table~\ref{tab:main_results}, \emph{Q/V LoRA + Mean} adapts query and value projections with temporal mean pooling. \emph{Q/K/V LoRA + Mean} also adapts key projections. The complete Audio Encoder combines Q/K/V adaptation and multi-scale pooling. It improves over both intermediate configurations and closely approaches text retrieval. The added LoRA and projection modules contain 20.19M parameters, excluding the pretrained backbone. The baseline text encoder contains approximately 568M parameters. These results support the effectiveness of our audio encoder design, which achieves near-baseline retrieval accuracy with a modest number of additional parameters.

\begin{table}[h]
    \caption{Clean retrieval on 1,000 queries. Intermediate rows compare different configurations. Scores are percentages.}
    \vspace{3pt}
    \label{tab:main_results}
    \centering
    \footnotesize
    \setlength{\tabcolsep}{2pt}
    \renewcommand{\arraystretch}{1.05}
    \begin{tabular*}{\columnwidth}{@{\extracolsep{\fill}}lccccc@{}}
        \toprule
        Method & R@1 & R@5 & R@10 & R@20 & MRR@20 \\
        \midrule
        ASR + Text Encoder
        & 67.7 & 84.3 & 89.2 & 92.1 & 75.2 \\
        Q/V LoRA + Mean
        & 59.7 & 78.7 & 84.6 & 89.1 & 68.5 \\
        Q/K/V LoRA + Mean
        & 63.0 & 81.2 & 85.7 & 89.9 & 71.4 \\
        \textbf{Audio Encoder}
        & 67.6 & 83.8 & 88.6 & 91.8 & 74.9 \\
        \bottomrule
    \end{tabular*}
    \vspace{-5pt}
\end{table}

\subsection{Latency Analysis}
\label{subsec:latency}

\begin{figure}[h]
    \centering
    \includegraphics[width=0.9\columnwidth]{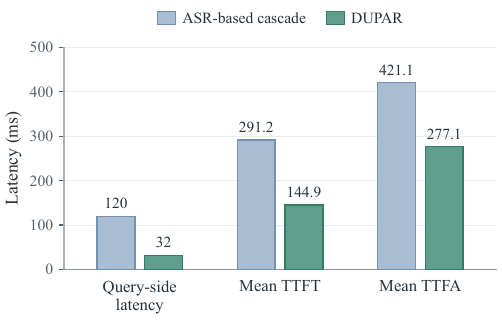}
    \caption{Pipeline latency comparison (lower is better). Query-side latency excludes generation, while TTFT and TTFA include generation up to the first output.}
    \label{fig:latency}
    \vspace{-15pt}
\end{figure}

Fig.~\ref{fig:latency} compares the speech-retrieval pipeline used in DUPAR with a conventional ASR-based cascade. The former achieves a $3.75\times$ query-side speedup and reduces mean TTFT and TTFA by 146.3 and 144.0\,ms, respectively. Both benchmarks use full-index retrieval.

\subsection{Noise Robustness}
\label{subsec:noise}

\begin{figure}[h]
    \centering
    \includegraphics[width=0.92\columnwidth]{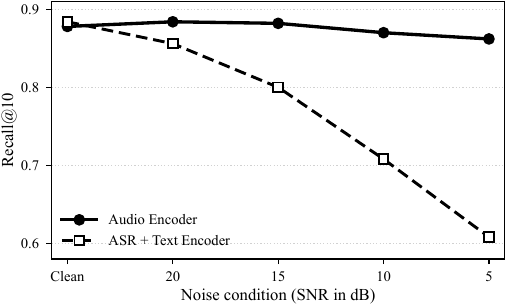}
    \caption{Recall@10 of Audio Encoder and ASR + Text Encoder under synthetic acoustic noise. Lower SNR indicates stronger noise.}
    \label{fig:noise}
    \vspace{-10pt}
\end{figure}

Across the five conditions in Fig.~\ref{fig:noise}, Audio Encoder averages 0.875 Recall@10, compared with 0.771 for ASR + Text Encoder. From clean speech to 5\,dB, their Recall@10 scores drop by 0.016 and 0.276, respectively. At 5\,dB, Audio Encoder achieves 0.862 versus 0.608, indicating greater robustness to the tested acoustic noise.

\subsection{Speaking-Style Analysis}
\label{subsec:breakdown}

\begin{table}[h]
    \vspace{-15pt}
    \caption{Retrieval across TTS speaking styles on a separate ASR-paired cohort. Scores are percentages.}
    \vspace{3pt}
    \label{tab:breakdown}
    \centering
    \footnotesize
    \setlength{\tabcolsep}{2pt}
    \renewcommand{\arraystretch}{1.05}
    \begin{tabular*}{\columnwidth}{@{\extracolsep{\fill}}lrcccc@{}}
        \toprule
        \multicolumn{1}{c}{\multirow{2}{*}{Speaking Style}}
        & \multicolumn{1}{c}{\multirow{2}{*}{$N$}}
        & \multicolumn{2}{c}{ASR + Text Encoder}
        & \multicolumn{2}{c}{Audio Encoder} \\
        \cmidrule(lr){3-4}\cmidrule(l){5-6}
        & & R@1 & R@10 & R@1 & R@10 \\
        \midrule
        Casual (Voice A) & 249 & 64.66 & 87.95 & 67.87 & 89.96 \\
        Casual (Voice B) & 258 & 67.83 & 88.76 & 72.48 & 91.09 \\
        Fast & 256 & 66.80 & 87.50 & 67.58 & 89.45 \\
        Sichuan & 237 & 62.03 & 85.65 & 70.46 & 89.03 \\
        \midrule
        Overall & 1,000 & 65.40 & 87.50
        & \textbf{69.60} & \textbf{89.90} \\
        \bottomrule
    \end{tabular*}
\end{table}

Table~\ref{tab:breakdown} compares the systems on the same waveforms. Audio Encoder improves overall Recall@1 by 4.20 percentage points. MRR@20 increases from 0.7318 to 0.7669. The largest Recall@1 gain occurs for Sichuan dialect. Recall@10 is higher across all four styles, although the Recall@1 difference for fast speech is not significant.

\subsection{Evidence Cache Analysis}
\label{subsec:evidence_cache}

We hold Audio Encoder fixed and build one-hop caches from the previous turn's actual full-index top-3. Under the conditions in Table~\ref{tab:evidence_cache}, restricting candidates improves Recall@1 by 32.26 percentage points. This supports local distractor suppression when graph connectivity and preceding retrieval provide relevant evidence.

\begin{table}[h]
    \vspace{-10pt}
    \caption{Conditional cache analysis on 31 follow-ups from 21 sessions. Adjacent gold chunks are graph-connected, and preceding actual top-3 retrieval contains the correct evidence. Scores are percentages.}
    \vspace{5pt}
    \label{tab:evidence_cache}
    \centering
    \footnotesize
    \setlength{\tabcolsep}{4pt}
    \renewcommand{\arraystretch}{1.05}
    \begin{tabular*}{0.8\columnwidth}{@{\extracolsep{\fill}}lcc@{}}
        \toprule
        Retrieval scope & R@1 & R@3 \\
        \midrule
        Full-index Retrieval & 35.48 & 54.84 \\
        Evidence-cache retrieval
        & \textbf{67.74} & \textbf{90.32} \\
        \bottomrule
    \end{tabular*}
    \vspace{-5pt}
\end{table}

\section{Conclusion}
\label{sec:conclusion}

We presented \textbf{DUPAR}, a dual-path conversational retrieval framework for knowledge-grounded voice assistants. Its fast path searches a cross-turn evidence cache, while a confidence-triggered slow path fuses full-index speech-query and ASR-transcript retrieval. On a domain-specific knowledge base, our speech retriever approaches the retrieval accuracy of ASR + Text Encoder with a $3.75\times$ query-side speedup. It also remains robust under synthetic noise and across speaking styles. Future work will evaluate the complete routing and cache-refresh loop in real multi-turn dialogues and broader acoustic conditions.

\vfill\pagebreak

\bibliographystyle{IEEEbib}
\bibliography{strings,refs}

\end{document}